\documentclass[usenatbib]{emulateapj}

\def\Msun{M_\odot}
\def\hMsun{h^{-1} \; M_\odot}
\def\hMpc{h^{-1} \; {\rm Mpc}}

\def\ergs{{\rm erg \; s^{-1}}}

\newcommand\geqsim{\lower.57ex\hbox{$\sim$}\llap{\raise.42ex\hbox{$>$}}$\,$}

\usepackage{graphicx}
\usepackage{epsfig,psfrag}
\usepackage{natbib}
\usepackage{pslatex}
\usepackage{multirow}

\shorttitle{Quasar MOF}
\shortauthors{Chatterjee et al.}

\begin{document}
\title{A Direct Measurement of the Mean Occupation Function of Quasars: Breaking Degeneracies between Halo Occupation Distribution Models} 
\author{Suchetana Chatterjee$^{1}$, My L.\ Nguyen$^{1}$, Adam D.\ Myers$^{1}$, Zheng Zheng$^{2}$ }
\affiliation{$^{1}${Department of Physics and Astronomy, University of Wyoming, Laramie, WY 82071.}\\ 
$^{2}${Department of Physics and Astronomy, University of Utah,
			Salt Lake City, UT 84112, USA}\\}
\email{schatte1@uwyo.edu}

\begin{abstract}

Recent work on quasar clustering suggests a degeneracy in the halo occupation distribution constrained from two-point correlation functions. To break this degeneracy, we make the first empirical measurement of the mean occupation function (MOF) of quasars at $z \sim 0.2$ by matching quasar positions with groups and clusters identified in the MaxBCG sample. We fit two models to the MOF, a power law and a 4-parameter model. The number distribution of quasars in host halos is close to Poisson, and the slopes of the MOF obtained from our best-fit models (for the power law case) favor a MOF that monotonically increases with halo mass. The best-fit slopes are $0.53 \pm 0.04$ and $1.03\pm 1.12$ for the power law model and the 4-parameter model, respectively. We measure the radial distribution of quasars within dark matter halos and find it to be adequately described by a power law with a slope $-2.3 \pm 0.4$. We measure the conditional luminosity function (CLF) of quasars and show that there is no evidence that quasar luminosity depends on host halo mass, similar to the inferences drawn from clustering measurements. We also measure the conditional black hole mass function (CMF) of our quasars. Although the results are consistent with no dependence on halo mass, we observe a slight indication of downsizing of the black hole mass function. The lack of halo mass dependence in the CLF and CMF shows that quasars residing in galaxy clusters have characteristic luminosity and black hole mass scales. 

\end{abstract}


\section{Introduction}

The number density and luminosity of quasars suggest that every massive galaxy, at some point, went through a quasar phase and is harboring a central supermassive black hole \citep[e.g.,][]{lyndenbell69,soltan82}. That such black holes have masses correlated with the velocity dispersion of the galactic bulge they inhabit implies a causal connection between galaxy evolution and black hole activity \citep[e.g.,][]{f&m00,gebhardtetal00, m&f01, tremaineetal02, gultekinetal09,grahametal11}. 

In addition, measurements of structure formation categorically demonstrate a simple relationship between the statistical distribution of galaxies and that of underlying dark matter halos \citep[e.g.,][]{w&f91, kauffmannetal93, nfw95, m&w96, kauffmannetal99, springeletal05a}. In tandem, these results suggest that the cosmological history of quasars is encoded in black-hole-mass to dark-matter-halo mass ($M_{{\rm BH}}$-- $M_{{\rm halo}}$) relationships \citep[e.g.,][]{ferrarese02}, which might arise as a combination of galaxy-mass to dark-matter-halo mass relationships and the quasar duty cycle \citep[e.g.][]{m&w01,hopkinsetal06a,c&w13}

The connection between black holes and their host dark matter halos has been mainly studied via clustering measurements of active galactic nuclei (AGN). Cosmological clustering is typically measured through the two-point correlation function \citep[2PCF; e.g.,][]{t&k69,arp70}. Under an assumed cosmology, the bias of an AGN \citep[the square root of the relative amplitude of AGN clustering to that of dark matter, e.g.,][]{kaiser84} can be inferred. By interpreting how dark matter halos of different mass are biased \citep[e.g.,][]{jing98,shethetal01}, a rough estimate of the typical mass of an AGN-hosting dark matter halo can be obtained.

Recently, a powerful analytic technique known as the halo occupation distribution \citep[HOD, e.g.,][] {m&f00, seljak00, b&w02, zhengetal05, z&w07} has started to be used to more fully interpret AGN clustering measurements \citep[e.g.,][]{wakeetal08, shenetal10, miyajietal11, starikovaetal11, allevatoetal11, richardsonetal12, k&o12, shenetal12a, richardsonetal13}. The HOD is characterized by the probability $P(N|M)$ that a halo of mass $M$ contains $N$ objects of a given type, coupled with the spatial and velocity distribution of the objects of interest inside their host halos.

The majority of AGN clustering measurements focus on luminous quasars---the centers of which are powered by highly accreting supermassive black holes---mostly because the extreme luminosity of quasars allows them to be used as a tracer of large-scale structure to very high redshift \citep[e.g.,][]{mortlocketal11}. Quasar clustering has been studied across a range of scales and redshifts \citep[e.g.,][]{croometal04, porcianietal04, croometal05, myersetal06, hennawietal06, myersetal07a, myersetal07b, hopkinsetal07d, coiletal07, shenetal07, daangelaetal08, shenetal09, rossetal09, padmanabhanetal09, hickoxetal11, shenetal12a, whiteetal12}. Recently two groups \citep{richardsonetal12, k&o12} performed a full HOD analysis of the 2PCF of quasars selected from the Sloan Digital Sky Survey (SDSS) and obtained constraints on the HOD properties of quasars at $z\sim1.4$.

These two recent measurements of quasar clustering adopted quite different HOD prescriptions. The \citet{richardsonetal12} work (R12 hereafter) used the parameterization of \citep[][C12 hereafter]{chatterjeeetal12}, which was derived from a study of low-luminosity AGN in a cosmological hydrodynamic simulation \citep{dimatteoetal08}. The mean occupation function (MOF hereafter) in C12 is modeled as a softened step function for the central component plus a rolling-off power law for the satellite component. \citet{k&o12} instead parameterized the MOF of central and satellite quasars such that they had the same shape, but with a different normalization---both follow a log-normal distribution, and the relative normalization is given by the mass-independent satellite fraction $f_{\mathrm sat}$.

\begin{figure}[t]
\begin{center}
\begin{tabular}{c}
\includegraphics[width=8cm]{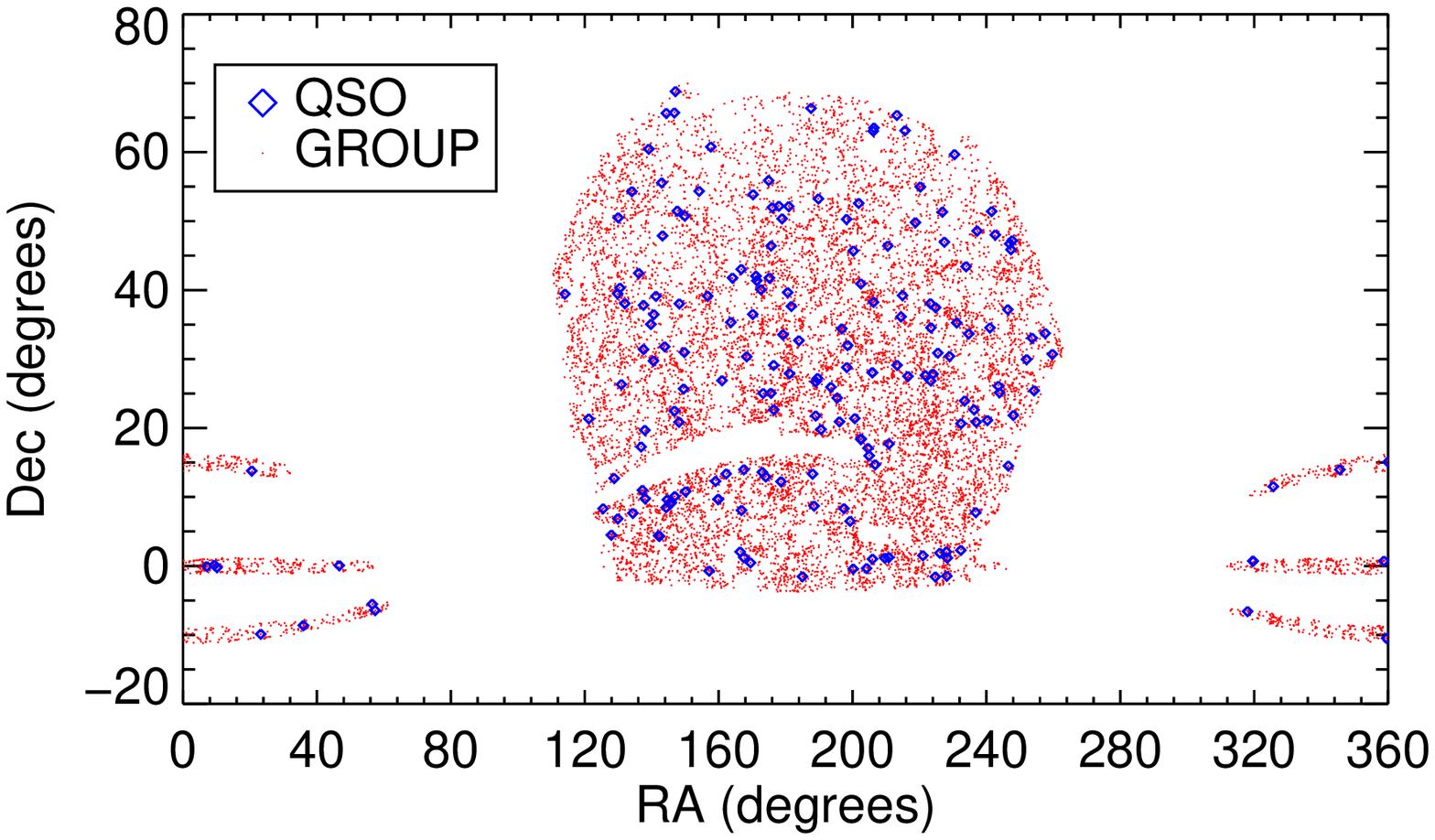}\\
\includegraphics[width=8cm]{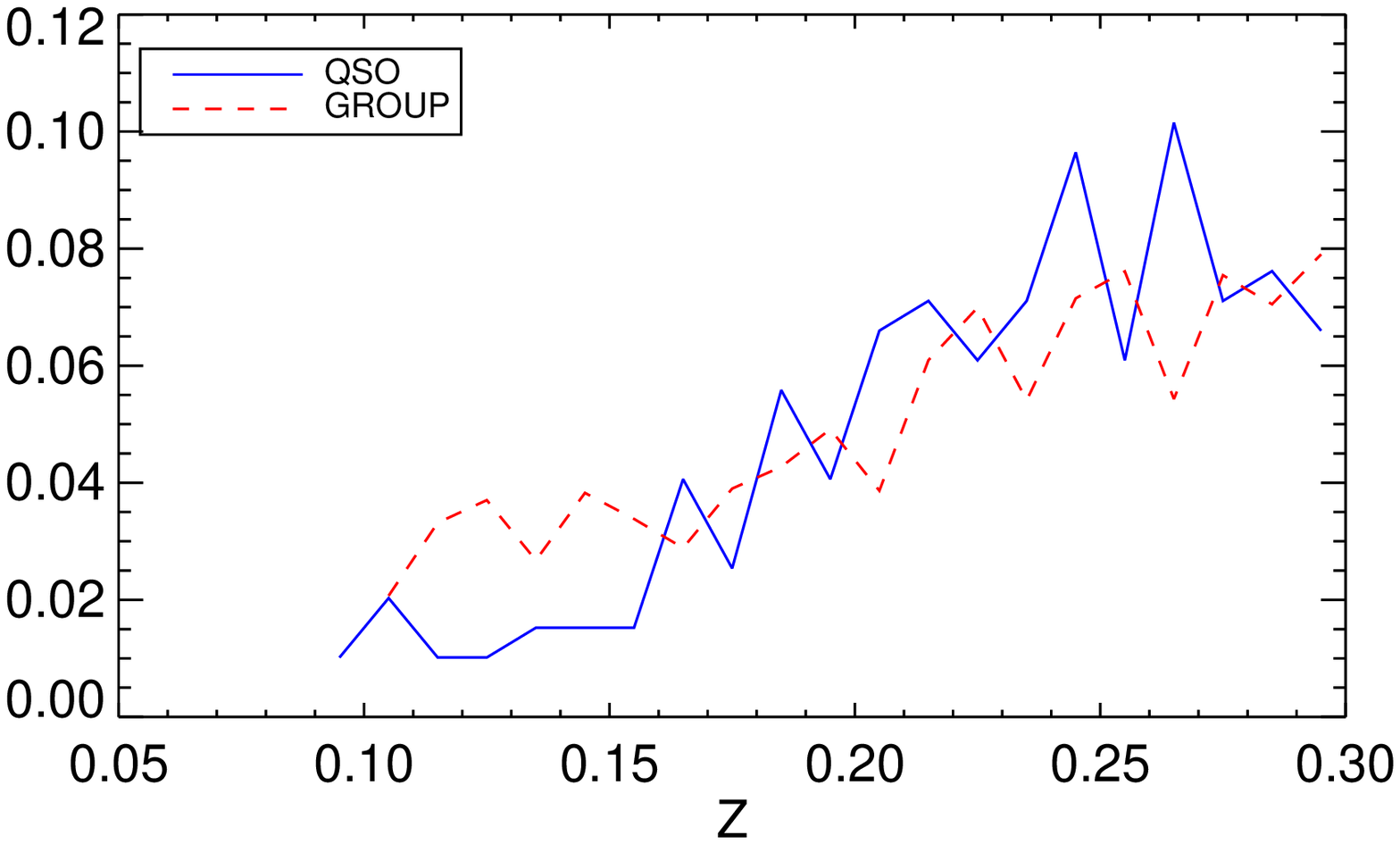}
\end{tabular}
 \caption{Positions and redshifts of quasars (in clusters) and the entire MaxBCG cluster sample. The top panel shows the positions of the quasars in host clusters (blue diamonds) and the red dots show the positions of MaxBCG clusters. The bottom panel shows the redshift distributions of quasars in clusters (solid blue line) and the entire MaxBCG sample (red dashed line). See \S 3 for a description of the methodology used to select a host group for each quasar. }
\end{center}
\end{figure}

Intriguingly, these two recent measurements of quasar clustering yielded quite different HOD parameters (e.g., shape of the MOF, satellite fraction), despite the similarity of the underlying data. It was noted in R12, and more recently in \citet{shenetal12a}, that different HOD models for quasars cannot be distinguished solely based on the 2PCF, and that additional observations beyond the 2PCF are needed to break HOD degeneracies. In this paper, we use SDSS quasar and galaxy group catalogs to empirically measure the MOF of quasars at $z \sim 0.2$. We then compare our findings with R12 ($z \sim 1.4$) and provide additional information on the HOD parameterization that can be fully exploited in future quasar surveys. Our approach of directly computing the MOF is analogous to the analysis conducted by \citet{allevatoetal12} for X-ray bright AGN.

Our paper is organized as follows. In $\S 2$, $\S 3$ and $\S 4$ we describe our data sets, outline our methodology, and present our results, respectively. We discuss and summarize our work in $\S 5$ and $\S 6$. Throughout the paper we assume a spatially flat, $\Lambda$CDM cosmology: $\Omega_{m}=0.28$, $\Omega_{\Lambda}=0.72$, and $h=0.71$ \citep{spergeletal07}. Unless otherwise stated, we quote all distances in comoving $\hMpc$ and masses in units of $\hMsun$. 

\section{Data}

We use data drawn from the Sloan Digital Sky Survey \citep[SDSS,][]{yorketal00}, which conducted 5 band ({\it ugriz}) photometry and extensive follow-up spectroscopy over more than $10{,}000$\,deg$^2$ of sky. Specifically, we use 1)\,the SDSS DR7 quasar catalog \citep{schneideretal10}, and 2)\,the MaxBCG cluster sample \citep{koesteretal07}, which we describe further in this section.

\subsection{Quasars}

The SDSS DR7 quasar catalog is described in detail in \citet{schneideretal10}. The catalog consists of $105{,}783$ spectroscopically confirmed quasars spanning a redshift range of $0.065 < z < 5.46$ and with an absolute $i$-band magnitude in the range $-30.28 \leq $M$_{i} \leq -22.0$. The catalog covers an area of $\sim 9830$ deg$^{2}$. Although the median redshift of the SDSS DR7 quasar catalog is $1.49$, we only consider quasars in the redshift range $0.1 < z < 0.3$ to match the redshift distribution of the MaxBCG galaxy clusters, which we discuss in the next section. 

\subsection{Galaxy Clusters}

\label{sec:maxbcg}

To trace the dark matter halos that host quasars we use the MaxBCG sample of galaxy groups, which is selected from SDSS imaging by combining brightest cluster galaxy (BCG) selection with the cluster red sequence method \citep[see][for more details]{koesteretal07}. MaxBCG group members are, essentially, all galaxies that lie within Rgal$_{200}$ of a likely BCG---provided that they are not more likely members of a second possible group. Rgal$_{200}$ is defined to be the radius within which the density of $-24 \leq {\rm M}_{r} \leq -16$ galaxies is at least $200\times$ the background . 

The MaxBCG catalog contains $13{,}823$ clusters with measured velocity dispersions larger than $\sim 400$ km/s. The sample is volume-limited and covers $\sim7500$\,deg$^2$ of sky in the redshift range $0.1 < z < 0.3$, with a median redshift of $0.25$. The photometric redshift error of the MaxBCG sample is $\sim0.01$ and is independent of redshift. The sample is $\sim 90$\% complete for clusters with masses greater than $\sim 10^{14}$M$_{\odot}$.

\begin{figure}
\begin{center}
\begin{tabular}{c}
\includegraphics[width=8.5cm]{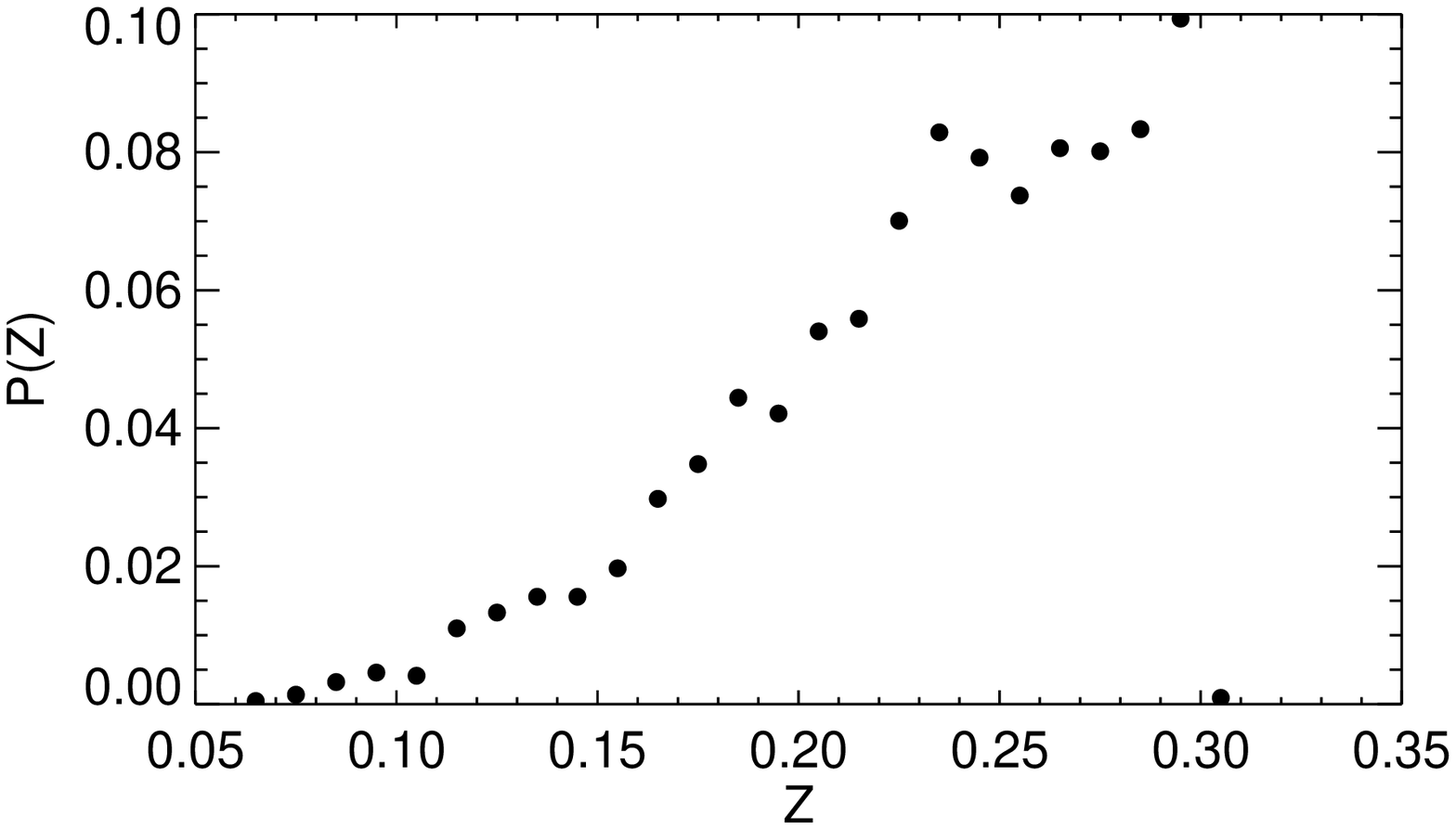}\\
\includegraphics[width=8.5cm]{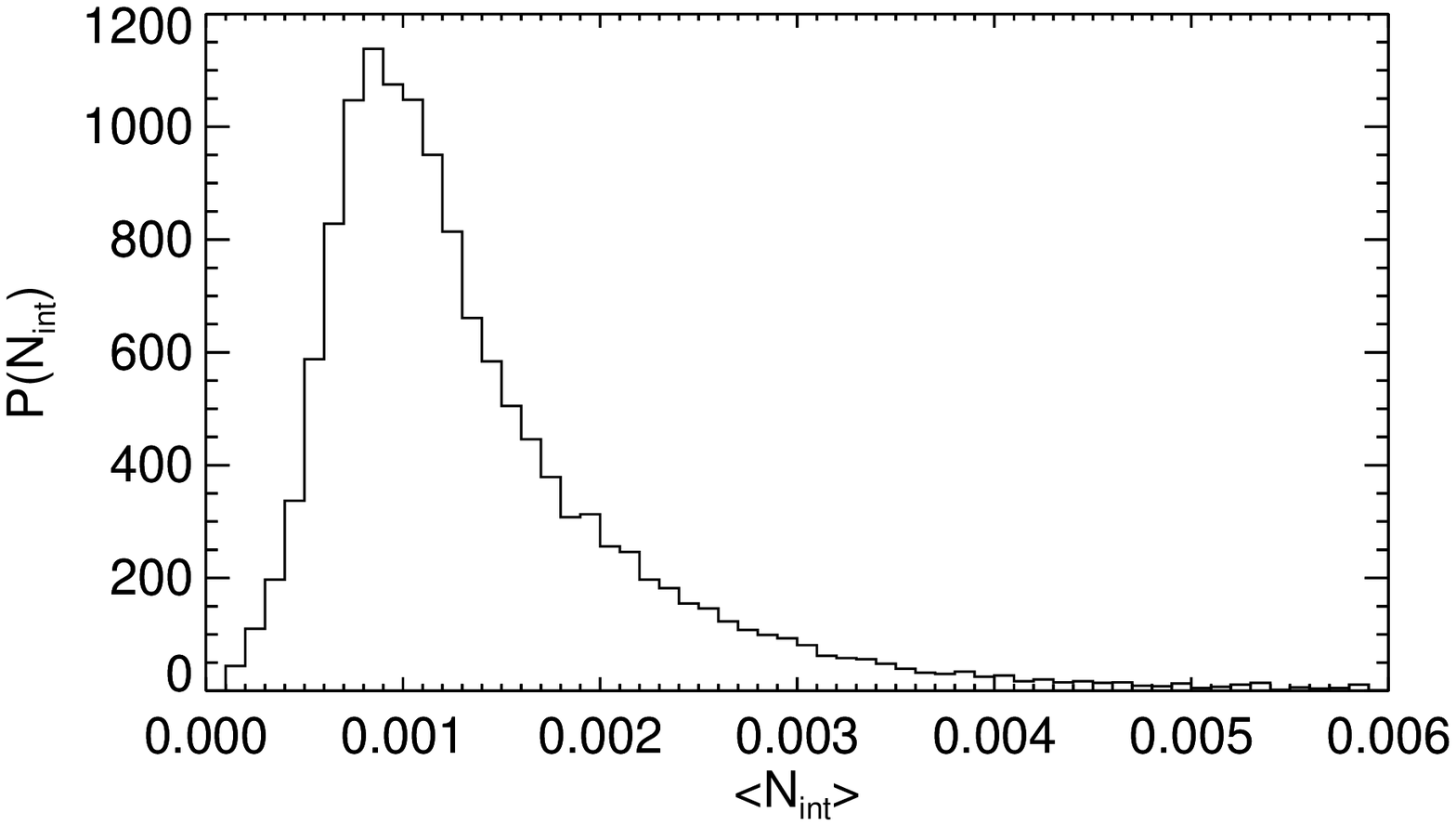}\\
\end{tabular}
 \caption{Possible contamination from interlopers. The top panel shows the redshift distribution of all quasars (within the MaxBCG redshift range) in the sample (black filled circles) and the dashed line depicts the best-fit power law, which we use to integrate Eq.\ 3. The bottom panel shows the distribution of the number of interlopers for all clusters in the sample.}
\end{center}
\end{figure}

\section{Methodology}
The first step in our methodology involves assigning host halos to our quasars. We follow an approach similar to the one that \citet{hoetal09} adopted in order to select host clusters for luminous red galaxies. We construct a cylindrical region around each galaxy cluster with a base radius $\theta_{200}$ and length $2\Delta z$, where $\theta_{200}$ is the angular equivalent of $R_{200}$ in the projected space and $\Delta z$ is the associated redshift interval.  Although $\Delta z$ is determined by the redshift error, we intend to set up an interval to safely select quasars that belong to the cluster. We define $R_{200}$ to be the radius at which the mean density of a cluster is $200$ times the mean matter density of the Universe. If the quasar falls within this cylindrical region then we identify the cluster as the quasar's host cluster. 

Due to fiber collisions, the SDSS has a limit ($55 \arcsec$) on the angular distance between targets on a single spectroscopic plate. Although most of the galaxies in the MaxBCG groups are defined purely from imaging, a small fraction could have been spectroscopically confirmed by the SDSS. However, the probability of such galaxies fiber-colliding with our quasar sample (which are all spectroscopically confirmed) is low, since quasars were given a higher priority than galaxies for targeting in SDSS-I/II \citep{blantonetal03}. In the case of two quasars within the same plate, there could be a possibility that we miss the spectroscopic observation of one quasar as a result of fiber collision with the other one, which would affect the occupation function and the radial distribution of quasars inside halos. This effect should be very small, but we discuss it further in section 4.2.

\begin{figure*}[t]
\begin{center}
\begin{tabular}{c}
\includegraphics[width=16cm]{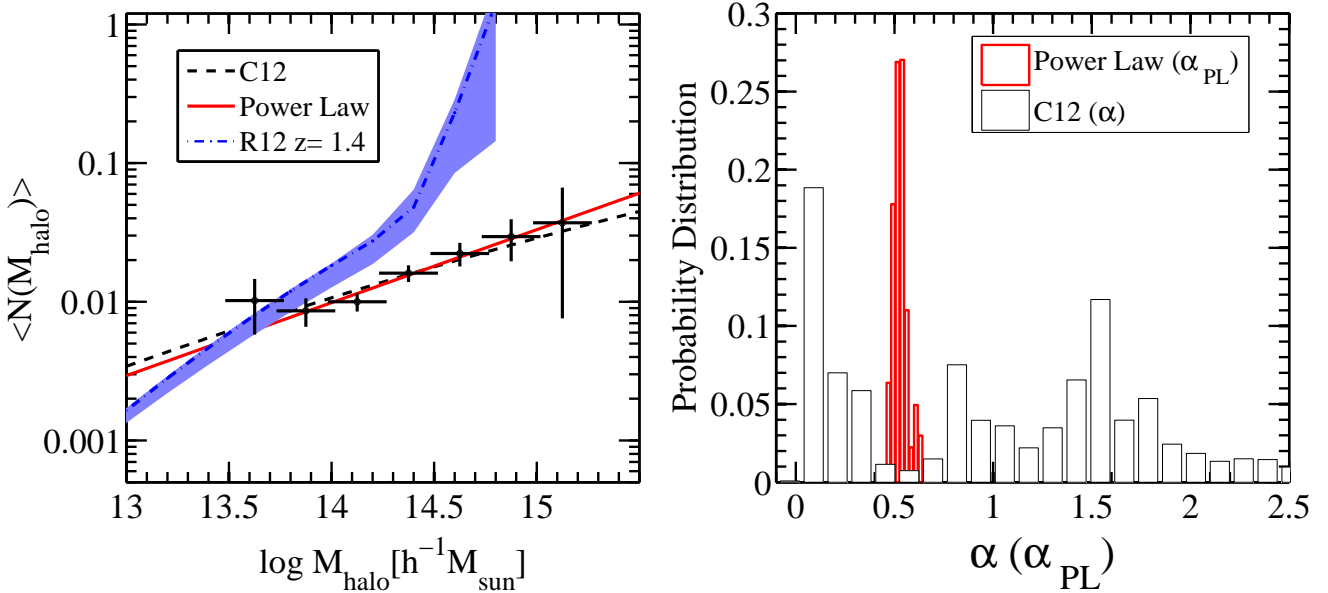}
\end{tabular}
 \caption{The mean occupation function (MOF) of quasars. In the right panel we show the distributions of the slopes of the occupation functions. The dashed black line is the best-fit theoretical curve corresponding to the C12 model (Eq.\ 4), and the red solid line represents the best-fit power-law model. The blue dot-dashed line represents the best-fit mean occupation function of R12 at $z=1.4$ obtained from HOD modeling of the 2PCF. The blue shaded region represents the $68\%$ confidence interval on the best-fit. The purple arrow refers to the mass scale beyond which we have $90\%$ completeness in cluster masses. In the right panel we show the probability distributions of our best-fit values for the slopes of the two models, which are $0.53\pm 0.04$, and $1.03\pm 1.12$ for the power-law model and the C12 model, respectively. Note that for the C12 model, the power-law slope ($\alpha$, Eq.\ 4) is actually the slope of the satellite occupation. For the power-law model the slope ($\alpha_{{\rm PL}}$) significantly favors a monotonically increasing occupation function with halo mass. The slope for the C12 model remains unconstrained.  }
\end{center}
\end{figure*}

We compute the cluster masses $M_{200}$ (i.e.\ the mass within $R_{200}$) using the modified optical richness estimates of \citet{rykoffetal12}
\begin{eqnarray}
   M_{200} &=& 5.58\times 10^{14}\left(\frac{N_{200}}{60}\right)^{1.08} h^{-1}M_{\odot}, \nonumber \\
   R_{200} &=& \left(\frac{3 M_{200}}{800\pi {\rm{\rho_{mean}}}}\right)^{1/3},
\end{eqnarray}
where $N_{200}$ is the number of galaxies within the cluster (the ``optical richness''). We adopt the mass definition corresponding to ``mean matter density ($\rho_{{\rm mean}}$)" as it is widely used in clustering measurements. We note that our measurements provide similar results whether we define $M_{200}$ in terms of the critical density of the Universe, or of the mean density. We verified this using the mass-richness relation involving critical density from \citet{rykoffetal12} and redefining our masses with respect to the critical density. From the mass estimates we obtain the radii ($R_{200}$ or $\theta_{200}$) of our cluster sample. 

We then apply both of the following criteria in order to identify quasar host groups:
\begin{eqnarray}
\theta \leq \theta_{200}  \nonumber \\ 
|z_{q} -z_{c}| & \leq & \Delta z,
\end{eqnarray} 
where $\theta$ is the angular separation between the quasar and the cluster center, and $z_{q}$ and $z_{c}$ are the redshifts of the quasar and the cluster respectively. We ensure unique hosts by assigning each quasar solely to the group to which it is closest.  In Fig.\ 1 we show the angular coordinates (top panel) and the redshift distribution (bottom panel) of quasars in clusters with reference to all the clusters in the MaxBCG sample. Note that the paucity of quasars in our sample at very low redshift ($0.1 \leq z \leq 0.15$) is not an effect of any cluster-related sample selection. Rather, it simply reflects a paucity of quasars in our parent sample \citep[c.f.\ Fig.\ 5 of][]{schneideretal10}

To choose an appropriate $\Delta z$ we adopt the following approach. We construct four mock quasar catalogs from our MaxBCG cluster sample using different underlying theoretical models of the MOF. The four different models are; the C12 central-only model, the model adopted by \citet{k&o12} and two different power-law models. We then use the MaxBCG group sample and the mock quasar samples to reconstruct the MOF using our methodology described in this section (above) with different choices of $\Delta z$. The redshift of the quasar with respect to the cluster redshift ( defined as $\Delta z_{{\rm qso}}$) is affected mainly by two components---the redshift error and the motion of quasars inside the cluster. 

The redshift error ($\Delta z_{{\rm err}}$) is a combination of the cluster redshift error and the quasar redshift error. The quasars are chosen spectroscopically and hence have very low redshift error ($\sim 0.001$). The error on our (photometric) cluster redshifts of $0.01$ \citep{koesteretal07} thus dominate the redshift error budget. The motion of quasars inside the cluster will cause a redshift difference between the quasar and cluster. The motion of quasars is proportional to $\sqrt{\rm GM/R}$, where M is the mass of the cluster and R is its size. With a simple assumption of a Maxwellian distribution with 1D velocity dispersion of $\sigma=\sqrt{\rm GM/(2R)}$ (i.e.\ an isothermal sphere) we can write $\Delta z_{{\rm qso}}=\sqrt{\Delta z_{{\rm err}}^{2}+\sigma^{2}}$. 

According to the above definition of velocity dispersion, the maximum velocity of quasars inside the cluster is $9.58\times10^{2}$km s$^{-1}$ ($\sigma \sim 0.003$, for a cluster of mass $1.8\times 10^{15}h^{-1}\Msun$). This implies $\sigma < \Delta z_{{\rm err}}$. Thus $\Delta z_{{\rm qso}} \sim \Delta z_{{\rm err}} \sim 0.01$. We adopt several possible $\Delta z$ cuts and compare our reconstructed MOF with the theoretical MOF model. We note that for all of the models, a choice of $\Delta z =0.03 (i.e., 3 \Delta z_{{\rm qso}})$ accurately recovers the true MOF. We thus adopt our fiducial value of $\Delta z =0.03$. However we note that our best-fit parameters for the MOF remain statistically identical for at least $0.01 < \Delta z < 0.03$. 

As discussed in \citet{hoetal09} there could be a finite probability of finding a quasar within a cluster just by chance (so-called ``interlopers''). This will potentially affect the occupation fraction of quasars. We thus applied a correction term to account for the interloper effect. For each cluster we calculate the possible number of interlopers and subtract that number from the occupation of quasars. We follow the procedure of \citet{hoetal09} to compute the number of interlopers via
\begin{equation}
\langle {\rm N_{int}} \rangle = \bar{n}\pi\theta_{200}^{2}\int_{z_{c}-\Delta z}^{ z_{c}+\Delta z}P(z_{q})dz_{q},
\end{equation}
where $P(z_{q})$ is the redshift distribution of quasars, $\theta_{200}$ is the size of the cluster, $\bar{n}$ is the mean surface density of quasars, $z_{c}$ is the redshift of the corresponding cluster, and $\Delta z =0.03$ for our purposes. The redshift distribution of all the quasars is shown in the top panel of Fig.\ 2. We estimate the quasar surface density to be $0.23$ deg$^{-2}$. The distribution of the number of interlopers ($\langle {\rm N_{int}} \rangle$) in each cluster is shown in the bottom panel of Fig.\ 2.

\begin{figure*}[t]
\begin{center}
\begin{tabular}{c}
\includegraphics[width=11.4cm, angle=-90]{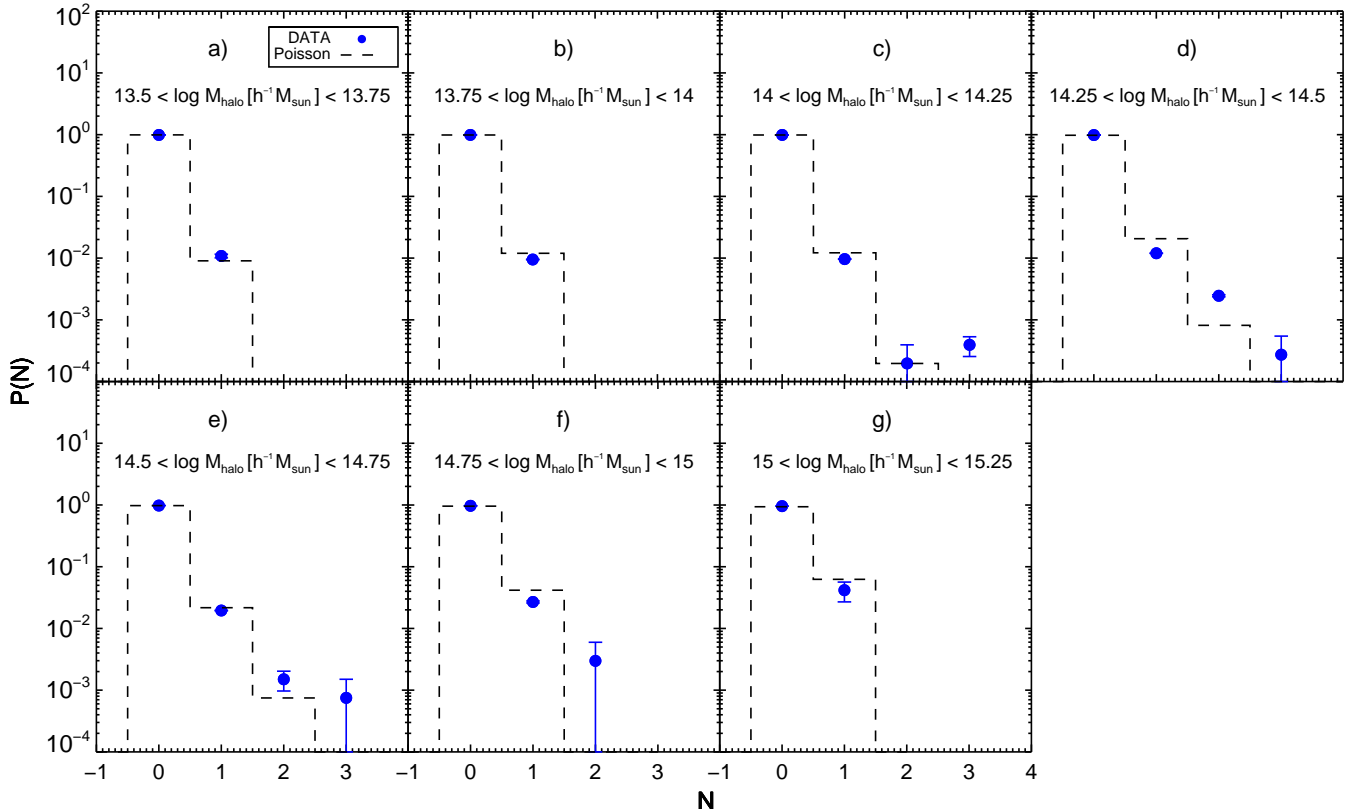}
\end{tabular}
 \caption{The normalized number distribution of quasars in the labelled host halo mass bins. The blue data points with error-bars show the actual distribution of quasars in halos and the black dashed histograms depict the theoretical Poisson distribution with a mean equal to the measured mean value of the occupation function corresponding to the particular mass bin (left panel of Fig.\ 3). The number zero represents the number of groups/clusters in the mass bin that do not host any quasar. The quasar number distribution seems to be close to a Poisson distribution, justifying the Poisson error-bars in Fig.\ 3. We note that the actual distribution will not be truly Poisson due to the presence of the central component (see the discussion in \S 4.1 and \citealt{zhengetal05} for more details).}
\end{center}
\end{figure*}

\section{Results}

From our host group and quasar catalogs we derive the MOF, the surface density profile of quasars within dark matter halos, and the conditional luminosity and black hole mass functions of quasars.  

\subsection{Mean Occupation Function}

The MOF for quasars is shown in the left panel of Fig.\ 3. The error on the mean number is assumed to be Poisson (Fig.\ 4 validates this assumption). The error in mass estimates is taken as $33\% $ in accordance with \citet{rykoffetal12}. We use the C12 model used by R12 and a simple power law (with a slope $\alpha_{{\rm PL}}$) to fit our MOF shown in Fig.\ 3. C12 has five free parameters and is given as a softened step function for the central quasars plus a rolling-off power law for the satellites,
\begin{eqnarray}
\langle N(M)\rangle &=& \frac{1}{2}\left[1+{\rm erf}\left(\frac{{\rm log} M-{\rm log} M_{\rm{min}}}{\sigma_{\rm{log M}}}\right)\right] \nonumber \\ & + & \left(\frac{M}{M_{1}}\right)^{\alpha} \exp \left(-\, \frac{M_{\mathrm{cut}}}{M} \right),
\end{eqnarray}
where $M_{\mathrm{min}}$ is the characteristic mass scale at which halos host, on average, 0.5 quasars; $\sigma_{{\rm log M}}$ is the characteristic transition width of the step function; $M_{1}$ is the approximate mass scale at which halos host, on average, one {\em satellite} quasar; $\alpha$ is the satellite power law index; and $M_{\mathrm{cut}}$ is the mass scale below which the satellite mean occupation decays exponentially. In the left panel of Fig.\ 3 we also overplot the best-fit occupation distribution (corresponding to the C12 model) of quasars at $z \sim 1.4$ from R12 (blue dot-dashed line).

We emphasize that the mass error will have a significant effect on the mean occupation function. Thus to obtain best-fit models we adopt a Monte Carlo approach.  We assume the errors on our cluster mass to be Gaussian-distributed and generate $20{,}000$ mock realizations of our original MaxBCG clusters. We measure the interloper-corrected MOF for these mock datasets. The interloper correction is incorporated in the following way. At each mass bin, we sum up
the interloper contribution from  all the clusters in the mass bin and then divide it by the total number of clusters corresponding to the same mass bin. We then minimize the $\chi^{2}$ for these $20{,}000$ simulated data pairs of a given mass and $\langle N(M)\rangle$. While performing the minimization we constrained our parameter space to the following limits: $10.0 \leq {\rm log}  M_{\mathrm{min}} \leq 25.0$, $10.0 \leq {\rm log} M_{1} \leq 25.0$. We fixed ${\rm log} M_{\mathrm{cut}} =10.0$ to reduce parameter degeneracies. This essentially assumes the satellite occupation to be a pure power-law, compared to the broken power-law model of Eq.\ 4. We do not apply any priors on $\sigma_{\rm{log M}}$ and $\alpha$.  

From the best-fit models of our simulated datasets, we quote the median and the standard deviation (from the mean) of the distribution of the $4$ parameters as our best-fits. We note that the mean in most cases is very close to the median but we preferred the median as the best-fit value since it is more effective in removing the outliers in the tail of the distribution. A similar approach was adopted for the power-law fits. In the right panel of Fig.\ 3 we show the distributions of the slope of the power law in each case. Note that for the C12 model, the power-law slope ($\alpha$) is actually the slope of the satellite occupation. The black thin solid histogram represents the distribution for the C12 model and the red thick solid histogram shows the distribution for the power-law fits. The best-fit slopes for the power law model and the C12 model are $ \alpha_{\rm PL} = 0.53 \pm 0.04$ and $\alpha = 1.03\pm 1.12$, respectively. We note that although the best-fit C12 model suggests a steeper slope (since the C12 slope is only for the occupation function of satellite quasars) it remains completely unconstrained. We also note that the C12 model prefers an occupation function where quasars are mostly identified as central quasars at these halo mass scales. We further discuss this issue in \S 5.1. 

In Fig.\ 4 we show the number distribution of the quasars in each host halo mass bin. The blue data points with error-bars represent the measured distributions. The black dashed histograms show the theoretical Poisson distribution, which was generated using the same number of quasars in each mass bin, assuming the mean occupation to be the theoretical mean of the Poisson distribution. The distribution of quasar numbers mimics the theoretical Poisson distribution, justifying our use of Poisson error-bars in Fig.\ 3. We note that only the satellite quasars would follow a Poisson distribution and the total distribution should truly be sub-Poisson \citep[e.g.,][]{kravtsovetal04, zhengetal05}.

\subsection{Radial Distribution}

Fig.\ 5 shows the radial distribution of the surface density of quasars within host halos. The profiles are normalized to the mean surface density of quasars ($\Sigma_{0}$) within $\theta_{200}$ \citep[e.g.,][]{n&k05}. The mean profile is obtained by stacking individual surface profile of quasars in each host halo and dividing the stacked profile by the total number of host halos. We fit a a simple power-law to the radial distribution. The error in the surface density estimate is assumed to be a quadratic combination of the Poisson errors from the number counts, and the systematic errors on the area estimate (assumed to be $22 \%$ which is two-thirds of the systematic error on the mass measurements. see Eq.\ 1). The error in the distance estimate is $\sim11\%$. We note that the error in the area dominates the error budget on the surface density estimate in most cases. We use a similar technique of generating mock simulations---as outlined in \S 4.1---to fit the radial profiles. 

As discussed in \S 3, fiber collisions will reduce quasar counts within $\sim55 \arcsec$ of another quasar target. This will potentially affect the surface density profiles in halos that contain multiple quasars on $\sim55 \arcsec$  scales. This should be a small effect, because in a stacked profile we should randomly sample the fiber-collided quasars. Nevertheless, we will conservatively ignore scales of $< 55 \arcsec$ when studying the radial distribution of quasars. Due to the variation in masses and redshifts of our cluster sample, $55 \arcsec$ will translate to different length scales. Thus for our stacked surface profile measurements we do not have a specific spatial scale beyond which we can ignore the fiber collision effect. We choose $0.25\theta_{200}$, roughly the fiber collision scale at the median redshift for the median mass ranges of our clusters, as the relevant scale below which we expect our measurements to be fiber-collision limited. The best-fit power-law (depicted by the black solid line in Fig.\ 5) slope is $-1.3 \pm 0.4$ when we exclude data points below the fiber collision scale. If we include all of the data points, the best-fit power-law (black dashed line in Fig.\ 5) slope becomes $-0.9 \pm 0.2$. Evidence from both theory and observations suggests that the radial distribution of black holes is closer to a power-law than an NFW \citep{nfw95} profile. We further discuss this issue in \S 5.2. Note that in fitting the surface density profile we assume that the interlopers are distributed uniformly within the cluster and there is no correlation between the positions of interlopers and their number distribution within clusters. If this is not true this can potentially affect the slope of the surface density profile.  

\begin{figure}
\begin{center}
\begin{tabular}{c}
\includegraphics[width=8cm]{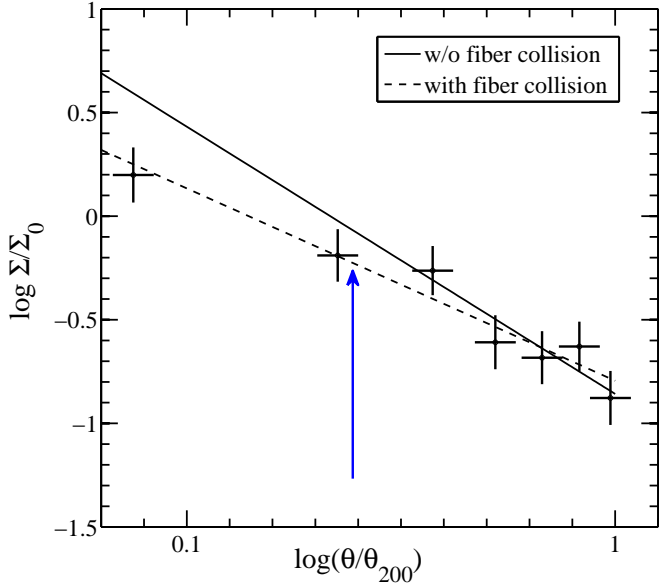}\\
\end{tabular}
 \caption{Surface density distribution of quasars. The black solid line represents the best-fit power law (slope of $-1.3 \pm 0.4$) describing the surface density profile. To obtain the best-fit power law slope, we excluded data points that lie below the fiber-collision scale. The black dashed line represents the best-fit power law (slope of $-0.9 \pm 0.2$) when the fiber collision scales are included. The error in the surface density is a combination of the Poisson error for number counts plus the error in area estimates (two-thirds of the mass error). The error in distance is assumed to be $11\%$. The blue arrow represents the physical scale corresponding to the fiber collision scale ($55 \arcsec$) for a typical cluster at the median redshift and having a mass equal to the median mass of the sample.  }
\end{center}
\end{figure}

\subsection{Conditional Luminosity and Black Hole Mass Functions}

The conditional luminosity function \citep[CLF, e.g.,][] {yangetal03}, for quasars is defined as the distribution of quasar luminosities $\Phi(L|M_{{\rm halo}})$ at a fixed halo mass $M_{{\rm halo}}$. The global luminosity function is given by 
\begin{equation}
\Phi(L) = \int \frac{dn}{dM_{{\rm halo}}}\Phi(L|M_{{\rm halo}})dM_{{\rm halo}},
\end{equation}
where $\Phi(L)$ is the quasar luminosity function and $dn/dM_{{\rm halo}}$ is the halo mass function. The CLF represents the differential form of the HOD, and can be used to investigate the luminosity evolution of quasar clustering. In Fig.\ 6 we show the CLF for our quasars, computed using $g$-band luminosities. We note that the CLF follows a log-normal distribution, similar to those observed by C12 for lower-luminosity AGN. The dashed line in each panel represents a reference log-normal curve which was used to fit the data in panel e of Fig.\ 6. Note that the best-fit log-normal curve (obtained from panel d of Fig.\ 6) adequately describes the distribution at all halo masses, indicating lack of luminosity evolution with host halo mass. The best-fit curve is described by
\begin{eqnarray}
\Phi(\log L) d\log L = \langle N(M_{{\rm halo}})\rangle P(\log L) d\log L, \nonumber \\ 
P(\log L) = \frac{1}{\sqrt{2\pi}\sigma_{{\rm log L}}}\exp\left[ -\frac{(\log L -\log L_c)^2}{2\sigma_{{\rm log L}}^2}\right]. 
\end{eqnarray}
The best-fit values for $\log L_{c}$ and $\sigma_{{\rm log} L}$ are $44.1 \pm 0.02$, and  $0.17 \pm 0.02$ respectively with luminosity in units of ergs s$^{-1}$. The quasar luminosities are computed using the K-corrections described in \citet{richardsetal06} assuming a power-law continuum slope of $-0.5$ in F$_{\nu}$. 

We also compute the conditional black hole mass function (CMF). The CMF $\chi(M_{{\rm BH}}|M_{{\rm halo}})$ is defined as the distribution of black hole masses at a fixed halo mass. The global mass function is 
\begin{equation}
\chi(M_{{\rm BH}}) = \int \frac{dn}{dM_{{\rm halo}}}\chi(M_{{\rm BH}}|M_{{\rm halo}})dM_{{\rm halo}},
\end{equation}
where $\chi(M_{BH})$ is the black hole mass function of quasars and $dn/dM_{{\rm halo}}$ is the halo mass function. The CMF is shown in Fig.\ 7. The black hole masses of our quasar sample have been taken from \citet{shenetal11}. The dashed line in each panel represents a reference log-normal curve which was used to fit the data in panel b of Fig.\ 7. The best-fit curve is described by
\begin{eqnarray}
\chi(\log M_{{\rm BH}}) d\log M_{{\rm BH}} = \langle N(M_{{\rm halo}})\rangle P(\log M_{{\rm BH}}) d\log M_{{\rm BH}}, \nonumber \\ 
P(\log M_{{\rm BH}}) = \frac{1}{\sqrt{2\pi}\sigma_{M}} \exp\left[ -\frac{(\log M_{{\rm BH}} -\log M_c)^2}{2\sigma_{M}^2}\right] \nonumber \\
 \end{eqnarray} 
where $\sigma_{M} = \sigma_{{\rm log M_{BH}}}$. The best-fit values for $\log M_{c}$ and $\sigma_{{\rm log M_{BH}}}$ are $8.1 \pm 0.1$, and  $0.57 \pm 0.08$ respectively with mass in units of h$^{-1}\Msun$. 
  
In Fig.\ 8 we plot the mean quasar luminosity/black hole mass (derived from the CLF/CMF) corresponding to each halo mass bin in the top/bottom panels respectively, with errors represented by 1 standard deviation (in log space). The red-dashed line in each panel represents the best-fit power-law model. The best-fit power-law slope for the top panel is $(-0.03 \pm 0.11$)---consistent with zero. This implies that, for the ranges of our samples, there is no significant correlation between quasar luminosity and host dark matter halo mass. The best-fit intercept value assuming a slope of zero (equivalent to a constant value) is $44.2\pm 0.1$ in units of logarithmic $ergs^{-1}$. Previous measurements of quasar clustering implied only a very weak luminosity dependence \citep[e.g.,][]{myersetal08, shenetal09, shenetal12a}---but here we establish this result independently, using empirical measurements of the mean occupation function and conditional luminosity functions. The best-fit power-law slope and an intercept value assuming zero slope (equivalent to a constant value) for the bottom panel of Fig.\ 8 are $(-0.15 \pm 0.25)$ and $(8.1 \pm 0.2)$, respectively, in logarithmic $\hMsun$ units. Although this does not represent a {\em significant} deviation from zero, the overall shape does not rule out the possibility of downsizing of black hole mass with host halo mass. We further discuss this result in \S 5.2. In each panel we overplot (black open circles) the luminosity and black hole mass of each individual quasar as a function of their host halo mass. 
\begin{figure*}
\begin{center}
\begin{tabular}{c}
\includegraphics[width=11.4cm, angle=-90]{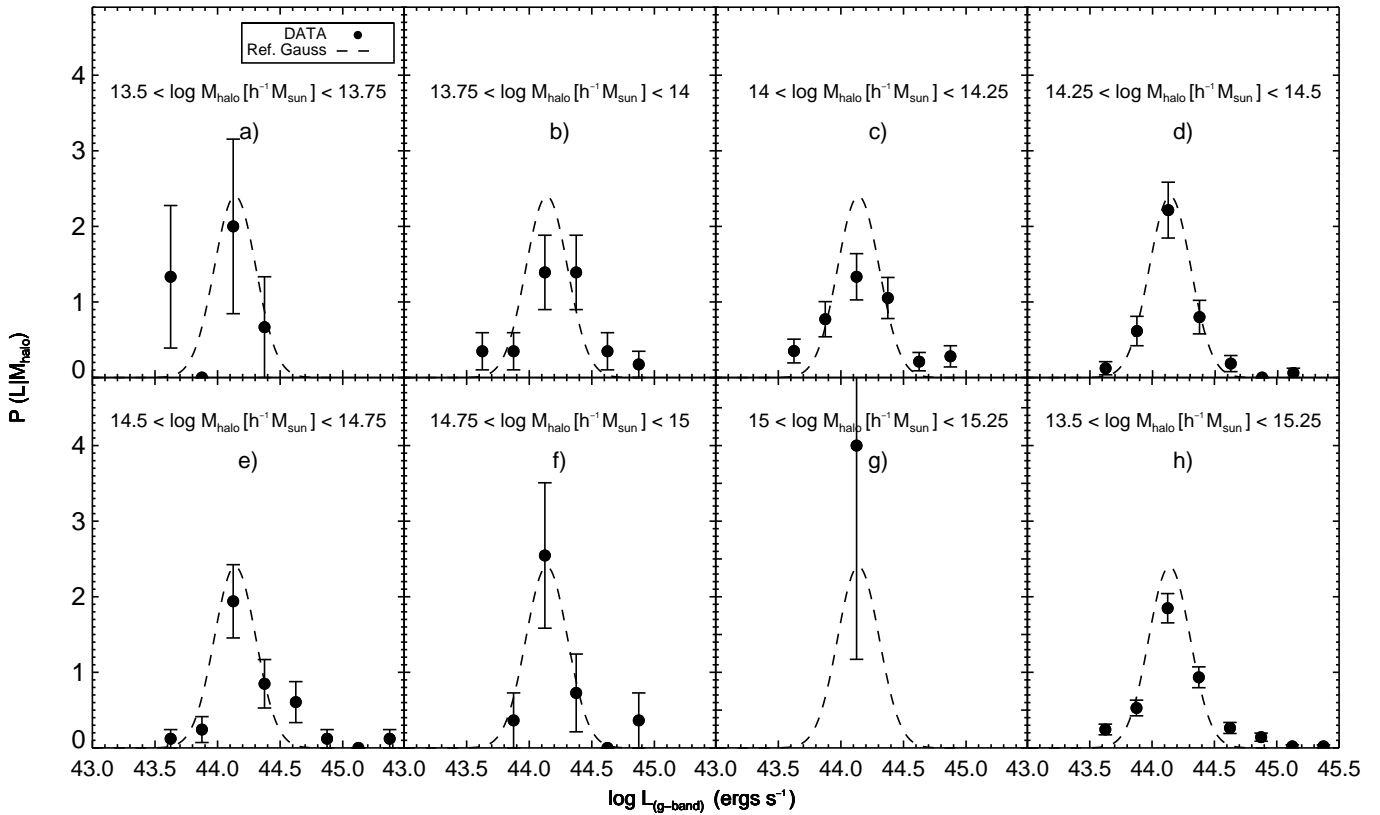}\\
\end{tabular}
 \caption{The conditional luminosity function (CLF) of quasars. The luminosities are $g$-band and errors are Poisson. The CLF follows a log-normal distribution, as expected. The luminosity distributions are identical in each halo mass bin indicating that quasar bias does not depend strongly on luminosity. The halo mass ranges corresponding to each bin are labeled. The black dashed line in each panel refers to the best-fit log-normal distribution corresponding to the data in panel e. The best-fit parameters are provided in Eq.\ 6. Panel h represents the luminosity distribution for the entire halo mass range which is essentially the global luminosity function of quasars in clusters at $z \sim 0.2$. }
\end{center}
\end{figure*}

\section{Discussion}

In this section, we discuss how systematic errors might affect our measurement of the MOF, and also compare our results with previous work. 

\subsection{Systematics}

To evaluate the effect of errors in the MaxBCG cluster masses on our MOF slopes we perform the following test. We assume the cluster mass measurements to be perfect and representative of the true masses of the clusters and then fit our MOFs without assuming any error on the mass distribution. Our best-fit slopes for the MOF (Fig.\ 3) are statistically identical (within $1 \sigma$ errors) with those obtained assuming $\Delta M/M = 0.33$. This is expected given that the mass errors are the same within each mass bin. The total $\chi^{2}$ values are $1.80$ and $1.81$ for the C12 and the power-law models, respectively. We note that the C12 model has degeneracies in its parameter space. To obtain a more meaningful comparison we thus fixed the satellite HOD parameters of C12 to their best-fit values from Fig.\ 3 and allowed the central HOD parameters to vary. In this scheme, we obtain a total $\chi^{2}$ of $1.98$ for the C12 central-only model. We thus conclude that the total MOF can be equally well fit by a C12 central-only model or a power-law model. 

In our analysis, we adopted photometric redshifts for the MaxBCG clusters from \citet{koesteretal07} which have a precision of $1\%$. To evaluate the effect of photometric redshifts on our measurements, we also compute our MOFs using the spectroscopic redshifts of the brightest cluster galaxy (BCG). We note that the results do not show any significant change when we use the spectroscopic BCG redshifts. As discussed before, we rely on Monte-Carlo techniques to adopt an appropriate $\Delta z$ for our measurements. 

We also evaluate the effect of changing the cluster radius on our MOF measurements. Our fiducial model assumes that a quasar will be identified as a member of the host group/cluster if the source lies within $R_{200}$ of the group/cluster. Our choice of $R_{200}$ is widely used in the literature and is motivated by, e.g., X-ray observations of galaxy clusters, for which $R_{200}$ is believed to coincide with the virial radius of the cluster \citep[e.g.,][]{leauthaudetal10}. Recomputing our MOFs using $0.5R_{200}$ or $2R_{200}$ barely affects our best-fit HOD parameters. Most notably, we obtain statistically identical power-law slopes with a slight difference in the normalizations.   

\begin{figure*}
\begin{center}
\begin{tabular}{c}
\includegraphics[width=11.4cm, angle=-90]{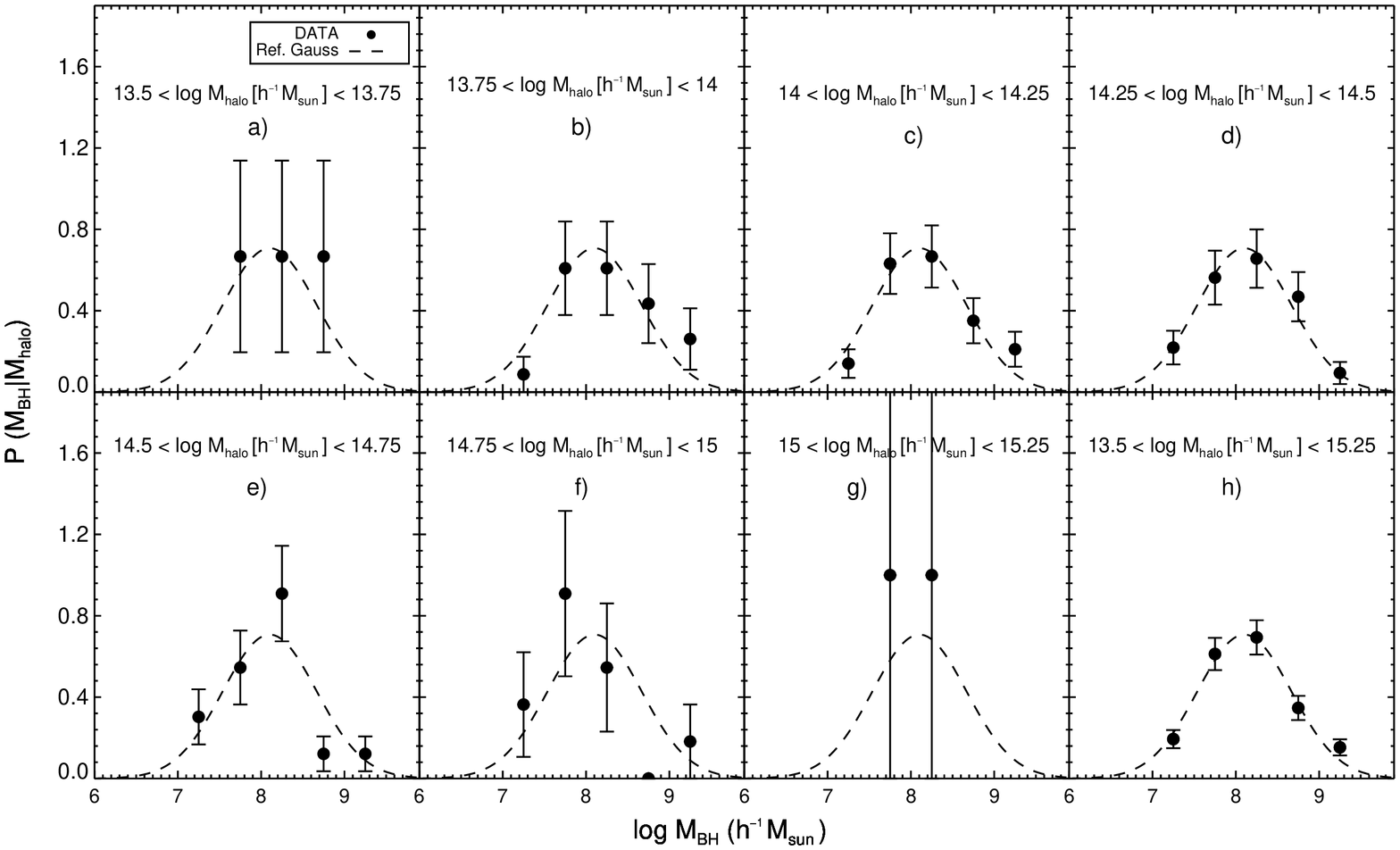}\\
\end{tabular}
 \caption{The conditional black hole mass function of the black holes that drive quasars. The black hole masses are computed from \citet{shenetal11} and the error-bars are Poisson. The halo mass ranges corresponding to each bin are labeled. The black dashed line in each panel refers to the best-fit log-normal distribution corresponding to the data in panel b. The best-fit parameters are provided in Eq.\ 8. Panel h represents the black hole mass distribution for the entire halo mass range, which is essentially the global black hole mass function of quasars in clusters at $z \sim 0.2$.  }
\end{center}
\end{figure*}

\subsection{Comparison with Other Works}

As discussed previously, we find that there exists a degeneracy in the HOD parameterization (particularly at the high mass end) while modeling the 2PCF. We emphasize that one of the major goals of our direct measurement is to break this degeneracy. R12 used the HOD model of C12 to fit the 2PCF of DR7 quasars and to recover the MOF (blue dot-dashed line in the left panel of Fig.\ 3). An alternative model by \citet{k&o12}, fit to similar data, found that the MOF decreased considerably more strongly with halo mass at the high mass end than found by R12. Although the measurements of the 2PCF and the subsequent HOD modeling by \citet{k&o12} and R12 were conducted at higher redshifts ($z\sim 1.4$) than for our work, we note that our low redshift empirical measurement favors the R12 model. Our best-fit power-law slopes for the MOF strongly favor ($\geqsim 10 \sigma$) a monotonically increasing occupation function with mass. To reconcile our results in this work with the negative MOF slope found by \citet{k&o12} would require strong redshift evolution in the MOF from $z \sim 0.2$ to $z \sim 1.4$.

To more completely compare measurements conducted using our method with the C12 model we would need to decompose the quasar occupation function into central and satellite components. To differentiate between central and satellite components we require information from the {\em member galaxies} in groups, rather than just the mean properties of the group. An enhanced version of the maxBCG catalog, known as the RedMapper catalog \citep{rykoffetal13} has been recently published, and includes information regarding the properties of member galaxies. We propose to conduct a central-satellite decomposition analysis, using the RedMapper catalog, in a future work. Exploiting additional information from the galaxy catalog will allow us to put additional constraints on the power-law slope of C12, which is currently unconstrained.   

Recently, \citet{allevatoetal12} used a novel approach to directly measure the occupation function of X-ray AGN in groups and clusters. They used a 6-parameter model to fit the data (see Eqs.\ 8 and 9 in \citealt{allevatoetal12}). The model is similar to C12, except for the asymptotic value of the central occupation function. We adopt unity for this asymptotic value in our work, whereas it is left as a free parameter in \citet{allevatoetal12}. The best-fit slope for the X-ray AGN sample recovered by \citet{allevatoetal12} is $0.22^{+0.41}_{-0.29}$ and $0.06^{+0.39}_{-0.28}$ for the C12 model and the power-law model, respectively. We note that we obtain a steeper slope than \citet{allevatoetal12}, but that their work probes different mass scales ($\log{\rm M_{halo}}$). The analysis of \citet{allevatoetal11} is restricted to $<14.16h^{-1}{\rm M_{\odot}}$, mass scales at which the contribution from satellite pairs to the MOF may well be smaller, which can affect the measurement of the satellite slopes.

In two recent studies \citet{degrafetal11b} and C12 measured the radial distribution of black holes (selected based on mass) and lower luminosity AGN (selected based on luminosity and host galaxy properties) in dark matter halos. These studies used cosmological hydrodynamic simulations that include black hole growth and feedback. C12 obtained a power law slope of $-2.33 \pm 0.08$ for AGN with bolometric luminosities $L_{\rm bol} \leq 10^{42} \; \ergs$ at $z \geq 1.0$. C12 also ruled out the NFW profile at $3\sigma$ for the same simulated AGN sample. As shown in Fig.\ 5, we fit the surface density profile rather than the 3D profile. Thus for the radial distribution our measurements suggest an equivalent best-fit power-law slope of $(-2.3 \pm 0.4)$, in excellent agreement with the value recovered by C12 from simulations. We emphasize that our quasar sample is substantially different (both in terms of luminosity and redshift) to the C12 sample and a direct comparison is not possible in this case. However the mass-selected sample of \citet{degrafetal11b} also suggests a power-law profile for the radial distribution. This general power-law form is believed to arise as a byproduct of black hole mergers \cite[]{degrafetal11a}.  Analytic models of dynamical friction in galaxy clusters also predict the AGN radial distribution to be steeper than NFW in the inner regions of the cluster \citep[e.g.,][]{nath08}. 

It is beyond the scope of this paper to quantify the effects of black hole mergers and/or dynamical friction on the radial distribution of quasars. But, a direct comparison of our technique to simulations is likely to require more precise measurements of the radial distribution from larger quasar and cluster samples. Our data potentially suffers from the fiber-collision effect at small scales and hence, without careful modeling of fiber collisions, we have insufficient information to test whether the radial distribution of quasars follows an NFW profile. Alternatively, a full analysis of cluster {\em members} using the RedMapper catalog \citep{rykoffetal13} would not suffer from fiber collisions.

\begin{figure}
\begin{center}
\begin{tabular}{c}
\includegraphics[width=8cm]{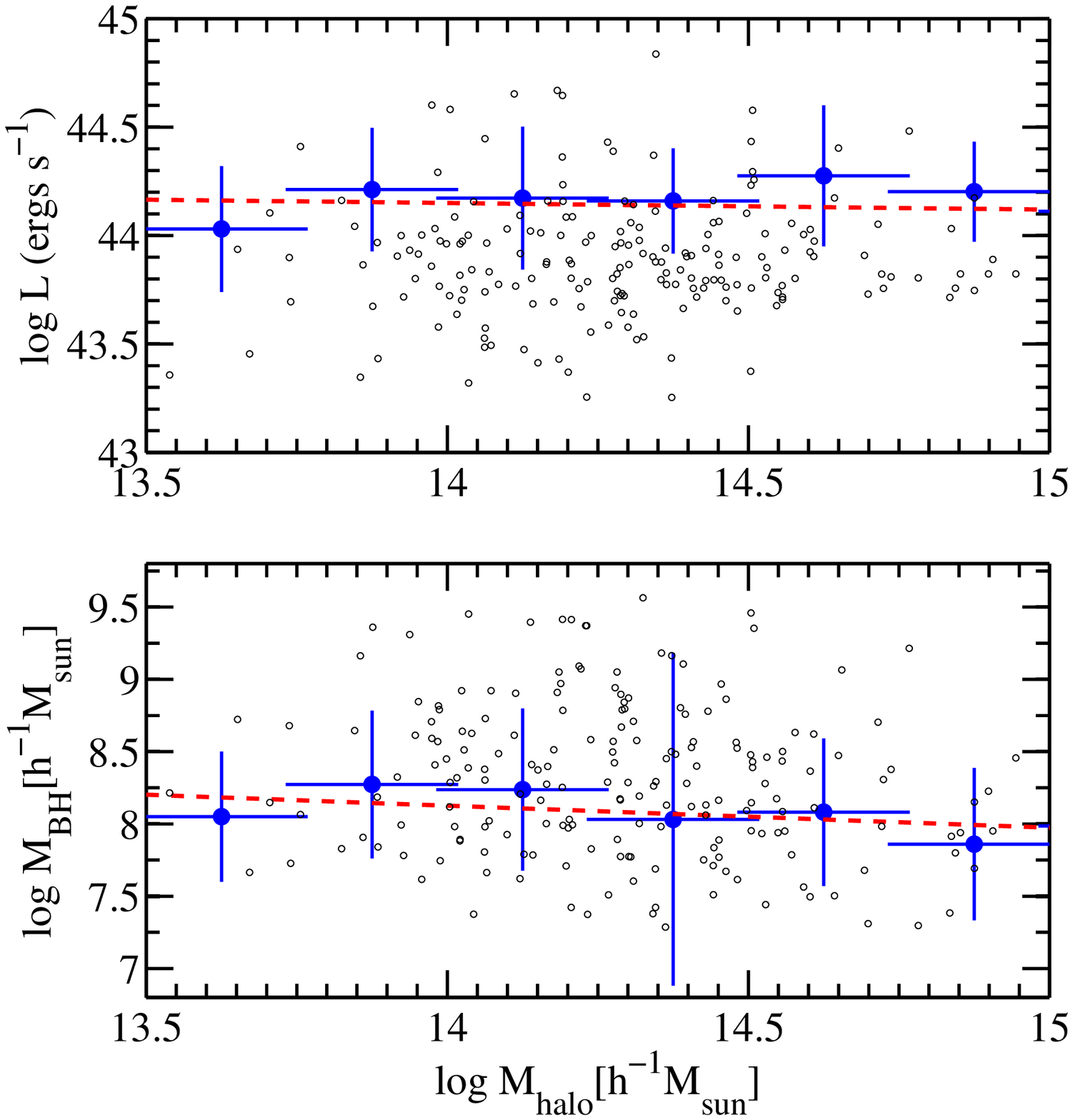}\\
\end{tabular}
 \caption{Top panel: The mean luminosity (computed from the CLF in Fig.\ 6) of quasars as a function of host halo mass. The error represents the spread of the distributions shown in Fig.\ 6. The red dashed line constitutes the best-fit power law. The slope of the power law ($-0.03 \pm 0.11$) is consistent with zero, showing that quasar bias does not depend strongly on luminosity. Previously a very weak luminosity dependence to quasar bias has been argued for based on clustering measurements. Here, we independently corroborate these results, for the first time, using empirical measurements of the CLF. Bottom panel: The mean masses (computed from the CMF in Fig.\ 7) of the black holes driving quasars as a function of host halo mass. Although we fit a simple power-law (red dashed line; $-0.15 \pm 0.25$) to our data, our results are not inconsistent with a broken power-law model (similar to the \citealt{c&w13} model), which would correspond to downsizing of the black hole mass function at high halo masses. However, we emphasize that our results are completely consistent with no dependence of the black hole mass function on host halos of quasars, similar to the inferences drawn from some clustering measurements \citep{shenetal09}. In each panel the open circles show the individual luminosity (top panel), and black hole mass (bottom panel) of quasars as a function of their host halo mass.}
\end{center}
\end{figure}

From HOD modeling of the 2PCF, R12 and \citet{k&o12} showed that the small-scale 2PCF prefers an NFW profile with a much higher concentration than typical dark matter halos at the redshifts they studied ($z\sim1.4$). In addition, our results are similar to the radial distribution of radio sources in clusters. \citet{l&m07} measured the radial profile of radio sources in clusters and showed that it is consistent with an NFW profile with a concentration of $25$, which is effectively equivalent to a power-law model. \citet{martinietal07} studied the radial distribution of X-ray selected AGN in clusters and found that AGN with X-ray luminosities above $10^{42} \ergs$ show stronger central concentrations than cluster host galaxies. Thus our results are in agreement with previous theoretical and observational studies.

The CLF and CMF measure how quasar luminosity and black hole mass are correlated with host dark matter halo mass. C12 and \citet{degrafetal11b} derived the CLF and CMF for low-luminosity AGN in simulations. As discussed before, the samples of C12 and \citet{degrafetal11b} are sufficiently different from our quasar samples that a direct comparison is inadvisable. However, there are several other analytic and numerical approaches that attempt to model these relationships \citep[e.g.,][]{s&o04, croton09,shen09,b&s09,shankaretal10,c&w13}.

\citet{c&w13} present a simple model for the relationship between quasars and their host dark matter halos in the redshift range $0.5 < z < 6$ using a linear relationship between black hole mass and host galaxy mass. The galaxy mass is connected to the halo mass through an empirically constrained relation.
Black holes shine at a fixed fraction of the Eddington luminosity during accretion episodes (equivalent to a light bulb model), and Eddington ratios ($\eta$) are drawn from a log-normal distribution that is independent of redshift. In Fig.\ 9 of their paper, \citet{c&w13} present the relationship between black hole mass and halo mass at different redshifts with $\eta$ held constant and with $\eta$ allowed to vary with redshift. The relationship is described by a broken power-law at all redshifts. At lower redshifts the power-law slope at the high-halo-mass end flattens. The lowest redshift for the \citet{c&w13} model is $0.5$, which is higher than the redshifts probed by our work in this paper. But, qualitatively, the \citet{c&w13} model predicts weaker dependence of black hole mass with halo mass at lower redshift. Although we do not detect any dependence in the CMF, our errors are too high to exclude weak dependences. We note that the weak evolution of CMF is also consistent with clustering measurements finding weak/no dependence on virial BH mass \citep{shenetal09}. However, an alternative possibility is that it is caused by the large uncertainty (about a factor of 3; \citealt{shen13}) in the mass measurements of the black holes, which tends to dilute dependence of the CMF on halo mass.

Under the assumption that quasar activity is triggered by major mergers in the hierarchical structure formation paradigm, \citet{shen09} derived scaling relations between quasars and their host dark matter halos. The black hole in this model follows an initial exponential growth at a constant Eddington
ratio of order unity until it reaches its peak luminosity, followed by a power-law decay \citep{shen09}. \citet{shen09} did not derive the redshift evolution of the halo mass-black hole mass slope and fix it at $\sim 1.6$. \citet{croton09} used the $M_{{\rm BH}}$--$\sigma$ relation and the quasar luminosity function to derive a scaling relation between black hole mass and quasar host halo mass, finding a slope that is equal to $1.39$ and that is independent of redshift. In many of these analytic models the relationship between halo mass and black hole mass has been derived from the $M_{{\rm BH}}$--$\sigma$ relation and the  $v_{{\rm c}}$--$\sigma$ relation, where $v_{{\rm c}}$ refers to the circular velocity of the bulge of the host galaxy \citep[e.g.,][]{m&f01,s&o04}. In general, this leads to an $M_{{\rm BH}}$--$M_{{\rm halo}}$ slope of 1.3--1.6. Critically, (at the low redshifts that we study), we find that the slope of the $M_{{\rm BH}}$--$M_{{\rm halo}}$ relation contradicts these models (bottom panel of Fig.\ 8).

Cosmological simulations of black hole evolution populate halos that cross a specified mass threshold with seed black holes of a given mass. These black holes then grow by accreting gas or by mergers \citep[e.g.,][]{dimatteoetal08, b&s09}. The slope of the black hole mass--halo mass scaling relation derived from these simulations depends on redshift and can be mostly described by a simple power law. The slope of this relationship lies in the range $\sim$ 1--1.5 \citep[e.g.,][]{c&d08, b&s10, degrafetal11a}. Our results are in tension with these simulations, although this may be because the luminosities that the simulations probe are several orders of magnitude below that of the quasars in our work.

We note that our observed lack of an $M_{{\rm BH}}-M_{{\rm halo}}$ correlation does not directly contradict the $M_{{\rm BH}}-\sigma$ relation. The RedMapper sample \citep[][which, as noted before, tracks the properties of MaxBCG member galaxies]{rykoffetal13} can be used to compute the $\sigma$ of cluster members. In future work, we intend to use this information to explicitly evaluate the $M_{{\rm BH}}-\sigma$ relation in the context of our non-evolving $M_{{\rm BH}}-M_{{\rm halo}}$ relation. It is important to note that we are looking at the $M_{{\rm BH}}-M_{{\rm halo}}$ relationship for luminous quasars residing in clusters of high halo mass and it is not confirmed if the $M_{{\rm BH}}-M_{{\rm halo}}$ relation persists on group-to-cluster scales \citep[e.g.,][]{mcconnelletal12}. Thus the lack of any correlation seen in our work does not necessarily contradict existing studies at lower halo mass and/or lower AGN luminosity.

\section{Summary}

In this work we employed an empirical measurement of the mean occupation function of quasars by identifying host halos of SDSS DR7 quasars from the MaxBCG group catalog. In the redshift range $0.1$--$0.3$ our measurements favor a monotonically increasing mean occupation function with halo mass. We fit a 4 parameter HOD model (C12) and a simple power-law model to our mean occupation function. The best-fit slopes are $0.53 \pm 0.04$, and $1.03\pm 1.12$ for the power-law model and the 4 parameter C12 model, respectively. Note that the slope for the C12 model refers to the satellite slope. We also show that the number distribution of quasars in dark matter halos is close to a Poisson distribution, as is observed for low-luminosity AGN in simulations. 

We obtain the radial distribution of quasars within dark matter halos and show that the measured surface profile is well described by a power-law model with a slope $-1.3 \pm 0.4$ (equivalent to $-2.3 \pm 0.4$ for the radial profile), in excellent agreement with the radial profiles of lower-luminosity AGN in cosmological simulations. We also measure the conditional luminosity function of quasars and show that it follows a log-normal distribution. From the conditional luminosity function we find no evidence of strong luminosity evolution of quasars with host halo mass, similar to the inferences drawn from clustering measurements of quasars at higher redshift. Finally, we compute the conditional black hole mass function of quasars and find no significant evidence that black hole mass is dependent on the dark matter halo hosting the quasar. 

Recent attempts to explain the quasar two point correlation function uncovered a large degeneracy in halo occupation models. Our empirical measurements provide an independent method with which to break this degeneracy, but are limited to the high-end of both the cluster-mass function and the quasar luminosity function. Our approach should thus become increasingly useful at higher redshift, over greater area, or using fainter samples of galaxies and/or quasars.

\section*{Acknowledgments}

We thank Yue Shen for discussions on estimates of black hole masses and for several other useful comments. We also thank James Bullock for his suggestion on pursuing central-satellite decomposition. Finally, we thank the referee for multiple constructive suggestions that improved the paper. SC and ADM were partially supported by the National Science Foundation through grant number 1211112, and by NASA through ADAP award NNX12AE38G, MLN was partially supported by NASA EPSCoR grant NNX11AM18A. ZZ is supported in part by NSF grant AST-1208891

\bibliography{mybib}{}
\bibliographystyle{apj}

\end{document}